\documentclass{article}
\usepackage{ijcai22}

\usepackage{times}
\usepackage{soul}
\usepackage{url}
\usepackage[hidelinks]{hyperref}
\usepackage[utf8]{inputenc}
\usepackage[small]{caption}
\usepackage{graphicx}
\usepackage{amsmath}
\usepackage{amsthm}
\usepackage{booktabs}
\usepackage{algorithm}
\usepackage{algorithmic}
\usepackage{tabls,threeparttable,tabulary}
\usepackage{multirow}
\usepackage{booktabs}
\usepackage{subfigure}
\usepackage{graphicx}
\usepackage{balance}
\usepackage{CJK}

\usepackage{textcomp}
\usepackage{xcolor}
\usepackage{amsmath}
\usepackage{algorithm}
\usepackage{algorithmic}

\floatname{algorithm}{Algorithm}

\usepackage{xspace}
\def\ie{\textit{i.e.}\xspace}

\def\eg{\textit{e.g.}\xspace}

\title{WaveFuzz: A Clean-Label Poisoning Attack to Protect Your Voice}

\author{
Yunjie Ge$^1$\and
Qian Wang$^1$\footnote{Qian Wang is the corresponding author.}\and
Jingfeng Zhang$^{2}$\and
Juntao Zhou$^1$\and
Yunzhu Zhang$^1$\and
Chao Shen$^3$\and
\affiliations
$^1$School of Cyber Science and Engineering, Wuhan University, Wuhan, Hubei, China\\
$^2$RIKEN Center for Advanced Intelligence Project, Tokyo, Japan\\
$^3$School of Cyber Science and Engineering, Xi'an Jiaotong University, Xi'an, Shaanxi,China\\
\emails
$\{$yunjiege, qianwang$\}$ @whu.edu.cn,
jingfeng.zhang@riken.jp,\\
$\{$ yunzhuzhang, juntaozhou$\}$ @whu.edu.cn,
chaoshen@mail.xjtu.edu.cn
}

\begin{document}

\maketitle

\begin{abstract}

People are not always receptive to their voice data being collected and misused. Training the audio intelligence systems needs these data to build useful features, but the cost for getting permissions or purchasing data is very high, which inevitably encourages hackers to collect these voice data without people's awareness.
To discourage the hackers from proactively collecting people's voice data, we are the first to propose a clean-label poisoning attack, called WaveFuzz, which can prevent intelligence audio models from building useful features from protected (poisoned) voice data but still preserve the semantic information to the humans. 
Specifically, WaveFuzz perturbs the voice data to cause Mel Frequency Cepstral Coefficients (MFCC) (typical representations of audio signals) to generate the poisoned frequency features. These poisoned features are then fed to audio prediction models, which degrades the performance of audio intelligence systems.
Empirically, we show the efficacy of WaveFuzz through attacking two representative types of intelligent audio systems, \ie, speaker recognition system (SR) and speech command recognition system (SCR). For example, the accuracies of models are declined by $19.78\%$  when only $10\%$ of the poisoned voice data is to fine-tune models, and  
the accuracies of models declined by $6.07\%$  when only $10\%$ of the training voice data is poisoned. 
 Consequently, WaveFuzz is an effective  technique that enables people to fight back to protect their own voice data, which sheds new light on ameliorating privacy issues. 

\end{abstract}

\section{Introduction}

\begin{figure}[t]
	\centering
	\includegraphics[width=1\linewidth]{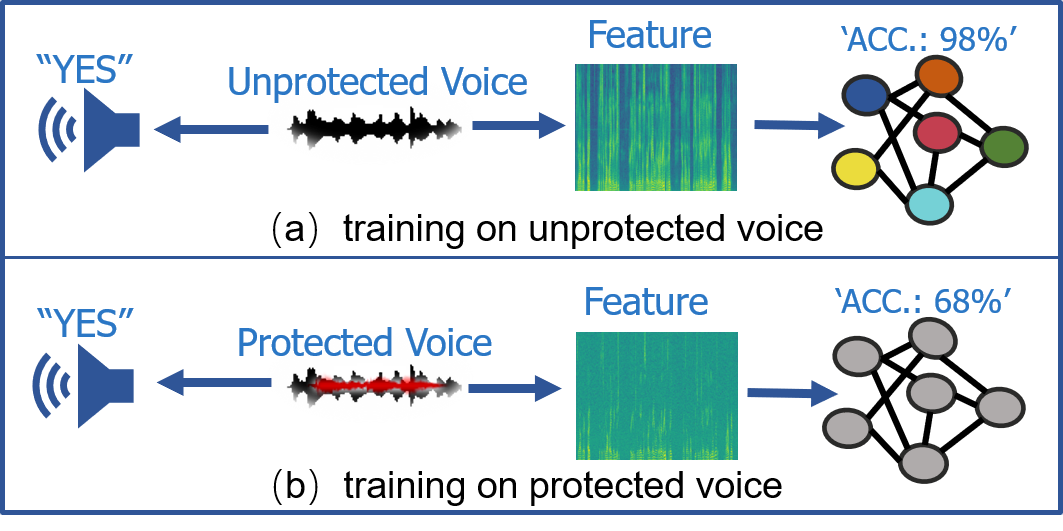}
	\caption{The protected voice example sounds like normal voice example but will produce the wrong features to the audio model, which degrades the accuracy of malicious models. }\label{fig:att}
\end{figure}

Due to the prevalence of social networks (\eg, Facebook, Twitter, and YouTube), individuals prefer to share their daily lives on the web.  For example, individuals interact with their friends and acquaintances with videos or audios in these nets. Therefore, such rich data at never-before-seen scales is contained on the web. Unfortunately, social networking sites always lack data protection, meaning that hackers can easily collect and misuse users' voice data as shown in Fig.~\ref{fig:att} (a). Nowadays, training high-performance audio intelligence systems requires massive voice data to build valuable features (\eg, voiceprint, pronunciation, and phoneme). Hackers prefer to collect these voice data without users' awareness from the web, \eg, using the crawler, owing to the expensive cost of getting permissions or purchasing voice data. However, people are not always receptive to their voice data being collected and used since voice data contains various information, including voiceprint and speech content.  Collecting and using people's voice data will bring privacy leakage concerns for the individual. Despite the great success in protecting user privacy (\eg, face)~\cite{DBLP:journals/scn/LiuZY17},  it is still an open challenge to mitigate privacy leaks about people's voice data.


To discourage the hacker from proactively collecting people's voice data, we propose  a clean-label poison attack, called WaveFuzz, which achieves the ability to generate indistinguishable protected voice data. Specifically, WaveFuzz elaborately perturbs voice data by maximizing the feature distance between unprotected (clean) voice data and protected (poisoned) voice data and minimizing their input distance, where features are extracted by MFCC (the de facto standard used by most intelligent audio systems). Then, the poisoned features extracted from poisoned voice data are fed into audio prediction models, where the  performance of the models will inevitably be degraded. Besides, we utilize a penalty term to limit the difference between benign and perturbed voice data to preserve the semantic meaning. Therefore, the protected voice data can transfer the correct information to humans but wrong knowledge to the machine. Moreover, the hacker 
is hard to use the poisoned samples to training modes and suffers from a punishment as shown in Fig.~\ref{fig:att} (b), \ie, performance decline .



We conduct experiments on two representative types of intelligent audio systems: SR and SCR. Moreover, we assume that the hacker may use users' voice data to fine-tune models or train models from scratch. For fine-tuning models, our WaveFuzz can  reduce the average of model accuracies by 19.78\%, 14.98\%, and 10.74\%  when 10\%, 5\%, and 0.5\% of the poisoned voice data is to fine-tune models. For training model from scratch,  our WaveFuzz can reduce the average of model accuracies by 6.07\%, 4.87\%, and 3.86\%  when 10\%, 5\%, and 0.5\% of  voice data is poisoned. Besides,  WaveFuzz outperforms the baseline method by 9.42\%, 6.81\%, and 4.79\% of accuracy decline, when 10\%, 5\%, and 0.5\% of the training voice data is poisoned. We also make a human study to validate that semantic meaning is preserved, where about 90\% volunteers can correctly understand the semantic meaning of poisoned voice data.

\section{Background and Related Work}
In this section, we introduce the components of intelligence audio recognition systems and then review related studies. 

\subsection{Background}
\begin{figure}[t]
	\centering
	\includegraphics[width=1.02\linewidth]{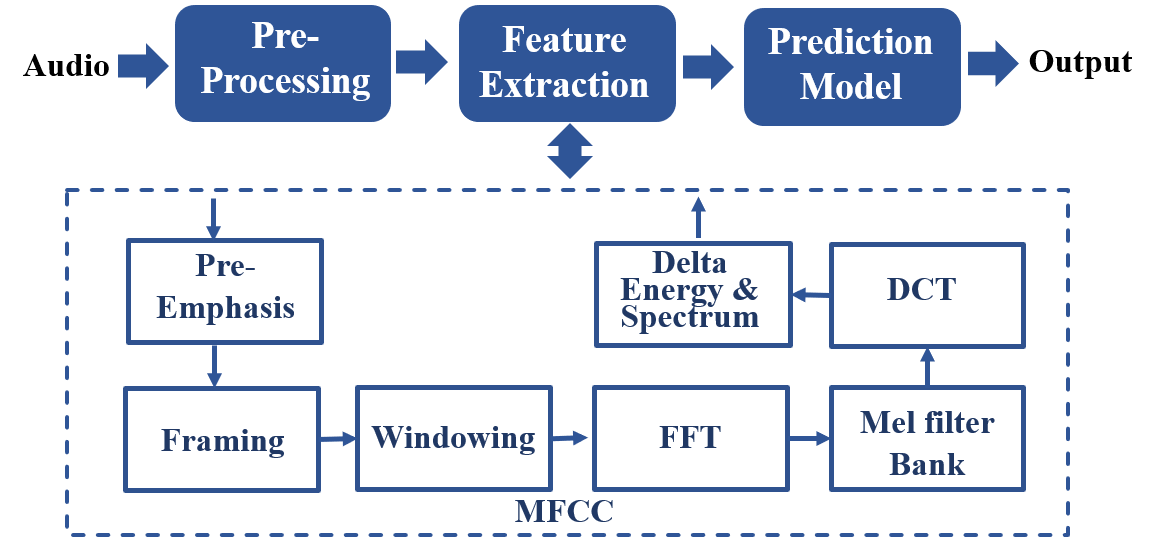}
	\vspace{-2em}
	\caption{The typical architecture of audio recognition systems and the overview of MFCC. }\label{fig:sr}
\end{figure}

The typical intelligence audio recognition system consists of three major components (as shown in Fig. \ref{fig:sr}): pre-processing, feature extraction, and the prediction model. Given a raw audio input, pre-processing will retain the frequencies in the range of human hearing and the segments higher than a specific energy threshold. Then, feature extraction will obtain the features from the processed audio by a feature extraction algorithm, such as Mel-frequency Cepstral Coefficients (MFCC)~\cite{muda2010voice}, Linear Predictive Coefficient (LPC)~\cite{itakura1975line}, and Perceptual Linear Predictive (PLP)~\cite{hermansky1990perceptual}. It is noted that MFCC is the de facto standard used in the audio field. Finally, the prediction model takes the extracted features as input and outputs the decisions.

MFCC~\cite{muda2010voice} can extract the parametric representation from an acoustic signal, which is in line with the acoustic characteristics of the human ear. As shown in Fig.~\ref{fig:sr}, MFCC consists of the following steps: 1) pre-emphasis increases the energy of signal at a higher frequency; 2) framing segments the speech samples into a small frame; 3) hamming windowing considers the next block and integrates all the closest frequency lines; 4) Fast Fourier transform (FFT) converting each frame from the time domain into the frequency domain; 5)  The energy spectrum is fed to a set of Mel-scale filter banks; 6)Discrete Cosine Transform (DCT) is used to convert the log Mel spectrum into the time domain; 7) Delta energy \& delta spectrum are used to add features related to the change in spectral features over time. 

\subsection{Related Work}
In this section, we simply review the related studies, including privacy protection methods and poisoning attacks.

\textbf{Privacy Protection:} To protect the users' privacy in the social networks, a bulk of studies~\cite{DBLP:conf/mva/DevauxPPC09,DBLP:conf/wacv/WilberSB16} utilized various obfuscation methods to blur human faces in a photo which can evade face detection. Moreover, several studies~\cite{DBLP:journals/corr/SzegedyZSBEGF13,DBLP:conf/iccv/OhFS17} used adversarial examples to protect photo privacy. The defender generates the adversarial example by adding invisible perturbation to the benign image, where the adversarial example can mislead the recognition models.

\textbf{Poisoning attack:}
The poisoning attacks can be divided into two classes~\cite{jagielski2018manipulating} based on the attack behaviors, including availability attacks and targeted attacks.
Availability attacks ~\cite{nelson2008exploiting,xiao2012adversarial} introduce failures in the final classifier by tampering with the training dataset, which have been the focus of many research efforts over time.
Targeted attacks~\cite{gu2017badnets,xiao2015feature,munoz2017towards} affect specific data points for the victim model. Recently, clean label poisoning attacks ~\cite{aghakhani2021bullseye,GeipingFHCT0G21,huang2020metapoison,zhu2019transferable} have been proposed.  The poisoned image example needs to be similar to the legal example but can perturb the model. Recent, VENOMAVE\cite{aghakhani2020venomave} is proposed to induce a targeted misclassification against automatic speech recognition (ASR) model using data poisoning. For example, the poisoned ASR model classifies the target frame $x_\tau$ as state $Z$, but performs well on the clean input. However, our paper aims to prevent unauthorized exploitation of voice data, a new aspect of data privacy and different from VENOMAVE. 

Despite great success in protecting the privacy of users' images,  there remains an urgent problem of preserving individual voices from being misused.



\begin{figure}
    \centering
    \includegraphics[width=\linewidth]{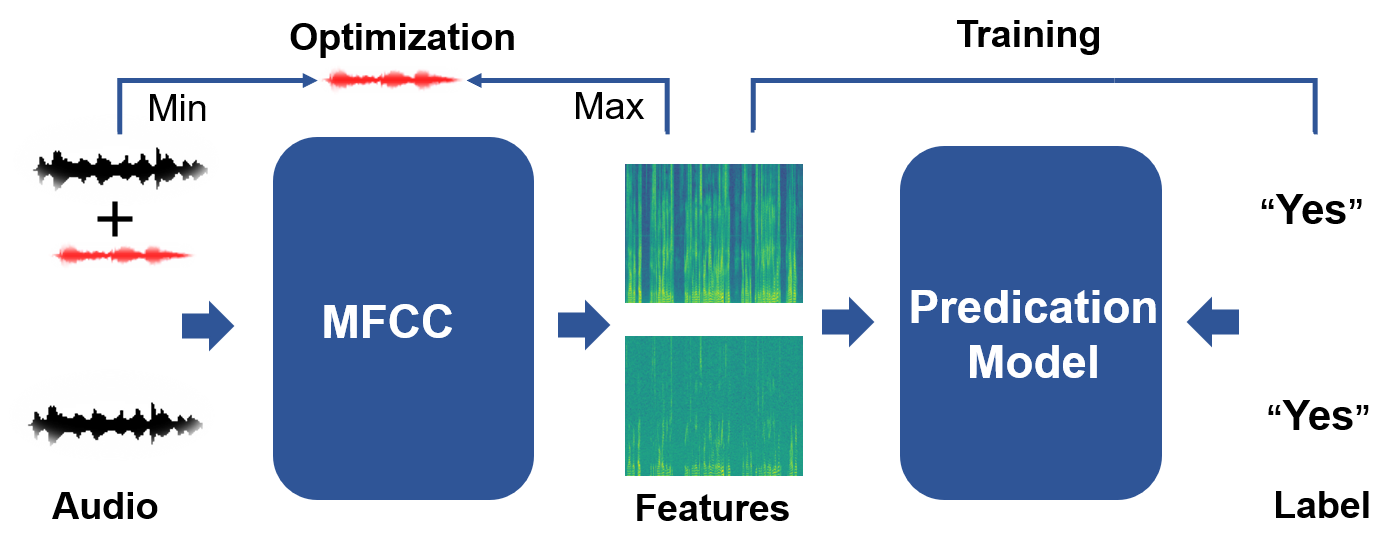}
    \vspace{-0.99em}
    \caption{The overview of WaveFuzz: The defender generates corrupted points by maximizing the distance of the MFCC features between normal and corrupted voice data, and minimizing their input distance. Then, poisoned features are fed into the prediction model will cause intelligence systems' performance to decline.}
    \label{fig:WF}
\end{figure}

\begin{algorithm}[!t]
	\caption{WaveFuzz}\label{AL:Wa}
	\begin{algorithmic}[1] 
		\REQUIRE {Normal example $x$; MFCC threshold $\varepsilon$;}
		\ENSURE {poisoned example $x^*$ }
		\STATE initialization: $ \delta^{(0)}$
		\STATE  $i=0$
		\WHILE{$\left \|\mathcal{MF} (x+\delta^{(i)})-\mathcal{MF} (x) \right \|_p < \varepsilon$} 
		\STATE $\delta^{(i)} \leftarrow E.q. ~\ref{eq:3}$
		\STATE $i= i+1$
\ENDWHILE
\STATE $x^* = x+\delta^{(i)}$
		\RETURN $x^*$
	\end{algorithmic}
\end{algorithm}

\section{Method}\label{se:me}
In the following, we introduce our WaveFuzz (see Fig.~\ref{fig:WF}), the first clean label poisoning attack to discourage hackers from collecting users' voice data.

\textbf{Insight.}
Given a voice example, the feature extraction component will extract the features as the input fed into the prediction model. Therefore, if the current features generated by the feature extraction algorithm are wrong, the wrong features will damage the prediction model inevitably. Moreover, the feature spectrum is not intuitive, making it difficult for humans to judge whether the spectrum is correct.

\textbf{Threat Model.}
We discourage the hacker from proactively collecting users' voice data by generating protected voice data. These protected voice examples are sound normal but can degrade the performance of the systems that want to steal the information from users' voice data.
We assume that the malicious systems use the MFCC to extract the features since MFCC is the most popular feature extraction algorithm in the audio field. Besides, we assume that the defender does not know the information of the prediction model.

\subsection{WaveFuzz}
Considering the popularity of MFCC, we use it as the feature extraction algorithm and denote it as $\mathcal{MF} (\cdot)$. MFCC will output features $\mathcal{MF}(x)$ for a voice example $x$, and the features are fed into the prediction model.

Given a voice input $x$ and the original label $y$, we want to craft a corrupted example $x^*$ that sounds like $x$, but $\mathcal{MF} (x^*)$ has a huge gap (distance) with $\mathcal{MF} (x)$. 
To achieve that, the overall solution can be divided into two steps. In the first step, we need to push the current example away from the original feature boundary. In other words, we need to find such example $x^*$ which meets the following constraint:
\begin{equation}
    \centering
    \begin{aligned}
        \left \| \mathcal{MF} (x^*)-\mathcal{MF} (x) \right \|_p> \varepsilon,
    \end{aligned}
\end{equation}
where $\varepsilon$ is a pre-defined threshold, and $\left \|  \cdot \right \|_p$ calculates the distance. We think that if the threshold $\varepsilon$ is large enough, the features $\mathcal{MF} (x^*)$ are away from the feature boundary of the current class .  

The current poisoned samples can compromise malicious models, where the models are hard to learn from the samples. However, the protected samples remain a weakness. That is, the individual may not understand the semantic content. To solve the problem, we make the poisoned example $x^*$  sound like the normal example $x$ in the second step. Therefore, the poisoned example $x^*$  should still satisfy the second constraint:
\begin{equation}
    \centering
    \begin{aligned}
    \left \|x^*-x \right \|_p < \epsilon, 
    \end{aligned}
\end{equation}
where $\epsilon$ is a threshold to limit the difference between the two examples. 

It is difficult to find such a example $x^*$ by solving the restricted problem. Therefore, we turn the restricted problem into the following optimization function:

\begin{equation}\label{eq:3}
    \centering
    \begin{aligned}
    \mathop{\arg\min}_{\delta }~-\left \| \mathcal{MF} (x+ \delta )-\mathcal{MF} (x) \right \|_p + \alpha \cdot \left \| \delta  \right \|_p,\\
    \end{aligned}
\end{equation}
where $\delta$ is the elaborate noise added to the original  example, and $\alpha$ is a parameter to balance the attack performance and stealthiness. Besides, we use the $L_2$-norm to calculate the distance between the two feature vectors.  

The overall pseudo-code is shown in Algorithm~\ref{AL:Wa}, which is easy to implement. We perform the optimization function (E.q.~\ref{eq:3}) until the distance between $\mathcal{MF}(x)$ and $\mathcal{MF}(x^*)$ is longer than the preselected threshold.

\begin{table}[t]
\centering
\small
\setlength{\tabcolsep}{1mm}{
\begin{tabular}{ccccc} 
\toprule
Task                                                         & \multicolumn{2}{c}{SR} & \multicolumn{2}{c}{SCR}                                                                                                    \\ 
\cmidrule(r){1-1}\cmidrule(lr){2-3}\cmidrule(lr){4-5}
Model                                                        & VggVox   & Resnet-18   & Attention                                             & CNN                                                          \\ 
\midrule
Dataset                                                      & Voxceleb & Voxceleb    & \begin{tabular}[c]{@{}c@{}}Speech \\Command V1\end{tabular} & \begin{tabular}[c]{@{}c@{}}Speech \\Command V2\end{tabular}  \\ 
\midrule
\begin{tabular}[c]{@{}c@{}}Feature \\Extraction\end{tabular} & FFT      & MFCC        & MFCC                                                        & FFT                                                        \\ 
\midrule
Accuracy                                                     & 82.20\%  & 79.07\%     & 93.78\%                                                     & 91.00\%                                                      \\
\bottomrule
\end{tabular}}
\caption{The information about the malicious models, the accuracies, and the datasets.}\label{ta:in}
\end{table}






\section{Experiment}
In this section, we consider that the hackers may use users' voice data to fine-tune pre-trained models (\eg, model updating) or train models from scratch. 

\subsection{Experiment Setup}
The information about the malicious models and the datasets is summarized in Table~\ref{ta:in}. We introduce the datasets and the relevant models used in our experiments in the following.
\begin{itemize}
    \item {\verb|Voxceleb|\footnote{https://www.robots.ox.ac.uk/~vgg/data/voxceleb/}}: It contains about 100,000 voices from 1,251 celebrities, extracted from interview videos uploaded to YouTube. VggVox is a speaker recognition model proposed by ~\cite{abs-1806-05622}.
   
    \item{\verb|SpeechCommand|\footnote{https://tensorflow.google.cn/datasets/catalog/speech\_commands}}: It contains 64,727 utterances from 1,881 speakers. Version 1 (V1) includes ten words. Version 2 (V2) adds four more command words. The detail information of the Attention-based model and  the convolutional neural networks (CNN) is in \cite{deandrade2018neural} and \cite{ArikKCHGFPC17}, respectively, where the two models are used to solve the speech command recognition task.
\end{itemize}

We quantify the following evaluation goals.
\begin{itemize}
    \item {\verb|DAcc.|}: DAcc. is defined as: $DAcc.= Acc.(\mathcal{F})-Acc.(\mathcal{F^*})$, where $ Acc.(\mathcal{F})$ is the accuracy of the model trained on clean dataset, and  $ Acc.(\mathcal{F})$  is the accuracy of the model trained on poisoned dataset. Besides, $Acc.(\mathcal{F})$ of all malicious models is reported in Table~\ref{ta:in}.
    
    \item {\verb|PR|}: Poison rate (PR) represents the rate of poisoned data in training data. PR is defined as: $PR= m/n$, where $m$ denotes the number of the poisoned examples, $n$ denotes the number of the benign examples.
    
    \item {\verb|SNR|}: We use SNR to describe the perturbation on the poisoned example $x^*$. $SNR=10\lg(\frac{x^*-x}{x})$, where $x$ denotes the original example.
    \end{itemize}
    
\subsection{Implementation and Training Details}
We generate malicious voice files following the formats of the datasets we use,  \eg, sampling rate and bit number. For fine-tuning, we directly use the poisoned voice data to fine-tune a pre-trained model. For training from scratch, we perturb a PR rate training samples. We will re-write the current samples using the corresponding poisoned samples. Given an voice sample, we use WaveFuzz to generate a compared poisoned sample and Adam optimizer~\cite{kingma2014adam} to optimize the sample. Besides, We utilize a substitute model to calculate a proper $\epsilon$ where the output is changed when the feature distance between the pure and poisoned samples is longer than the $\epsilon$.

\begin{table}[t]
\resizebox{0.94\linewidth}{!}{
\centering
\begin{tabular}{cccccc} 
\toprule
Model                                                                     & PR(\%)  & DAcc.(\%)     & SNR    & $\epsilon$            & $\alpha$                \\

\midrule
\multicolumn{6}{c}{TASK:SR}                  \\ 
\midrule
\multirow{4}{*}{\begin{tabular}[c]{@{}c@{}}VggVox\\Model\end{tabular}} & 10  & 30.34 & 11.13 & \multirow{4}{*}{400} & \multirow{4}{*}{0.1}  \\
                                                                          & 5    & 24.55 & 11.15 &                       &                         \\
                                                                          & 1    & 22.48 & 11.19 &                       &                         \\
                                                                           & 0.5    & 20.31 & 11.25 &                       &                         \\

\midrule      
\multicolumn{6}{c}{TASK:SCR}                  \\ 
\midrule

\multirow{4}{*}{\begin{tabular}[c]{@{}c@{}}Attention\\Model\end{tabular}} & 10   & 14.63 & 4.14 & \multirow{4}{*}{300} & \multirow{4}{*}{0.1}  \\
                                                                          & 5    & 10.64 & 4.92 &                       &                         \\
                                                                          & 1    & 7.42 & 5.40 &                       &                         \\
                                                                          & 0.50 & 6.18 & 5.12 &                       &                         \\ 
\midrule
\multirow{4}{*}{\begin{tabular}[c]{@{}c@{}}CNN\\Model\end{tabular}} & 10   & 14.37 & 7.89 & \multirow{4}{*}{1700} & \multirow{4}{*}{0.1}  \\
                                                                          & 5    &9.75 & 7.80 &                       &                         \\
                                                                          & 1    & 7.37 & 7.73 &                       &                         \\
                                                                          & 0.50 & 5.75 & 7.69 &                       &                         \\ 
                 
\bottomrule
\end{tabular}}

\caption{WaveFuzz against models that are fine-tuned by the poisoned voice data.}\label{TA:PE1}
\end{table}

\begin{table}[t]
\centering
\resizebox{0.94\linewidth}{!}{
\begin{tabular}{cccccc} 
\toprule
Model                                                                     & PR(\%)  & DAcc.(\%)     & SNR    & $\epsilon$            & $\alpha$                \\

\midrule
\multicolumn{6}{c}{TASK:SR}                  \\ 
\midrule
\multirow{4}{*}{\begin{tabular}[c]{@{}c@{}}VggVox\\Model\end{tabular}} & 10  & 7.28 & 10.14 & \multirow{4}{*}{1000} & \multirow{4}{*}{0.1}  \\
                                                                          & 5    & 5.23 & 10.28 &                       &                         \\
                                                                          & 1    & 4.00 & 10.04 &                       &                         \\
                                                                          & 0.5    & 2.15 & 10.27 &                       &                         \\

\midrule      
\multirow{4}{*}{\begin{tabular}[c]{@{}c@{}}Resnet-18\\Model\end{tabular}} & 10  & 2.48 & 9.96 & \multirow{4}{*}{1000} & \multirow{4}{*}{0.1}  \\
                                                                          & 5    & 1.84 & 10.14 &                       &                         \\
                                                                          & 1    & 1.54 & 10.23 &                       &                         \\
                                                                          & 0.5    & 3.39 & 10.05 &                       &                         \\

\midrule      

\multicolumn{6}{c}{TASK:SCR}                  \\ 
\midrule

\multirow{4}{*}{\begin{tabular}[c]{@{}c@{}}Attention\\Model\end{tabular}} & 10   & 7.89 & 3.88 & \multirow{4}{*}{1700} & \multirow{4}{*}{0.1}  \\
                                                                          & 5    & 6.97 & 3.90 &                       &                         \\
                                                                          & 1    & 6.31 & 3.83 &                       &                         \\
                                                                          & 0.50 & 4.68 & 3.91 &                       &                         \\ 
\midrule
\multirow{4}{*}{\begin{tabular}[c]{@{}c@{}}CNN\\Model\end{tabular}} & 10   & 6.62 & 6.89 & \multirow{4}{*}{1700} & \multirow{4}{*}{0.1}  \\
                                                                          & 5    &5.45 & 7.45 &                       &                         \\
                                                                          & 1    & 5.35 & 7.26 &                       &                         \\
                                                                          & 0.50 & 5.23 & 7.76 &                       &                         \\ 
                 
\bottomrule
\end{tabular}}

\caption{WaveFuzz against models that are trained from scratch by the poisoned voice data.}\label{TA:PE2}
\end{table}

 \begin{figure}[t]
	\centering
	\includegraphics[width=0.90\linewidth]{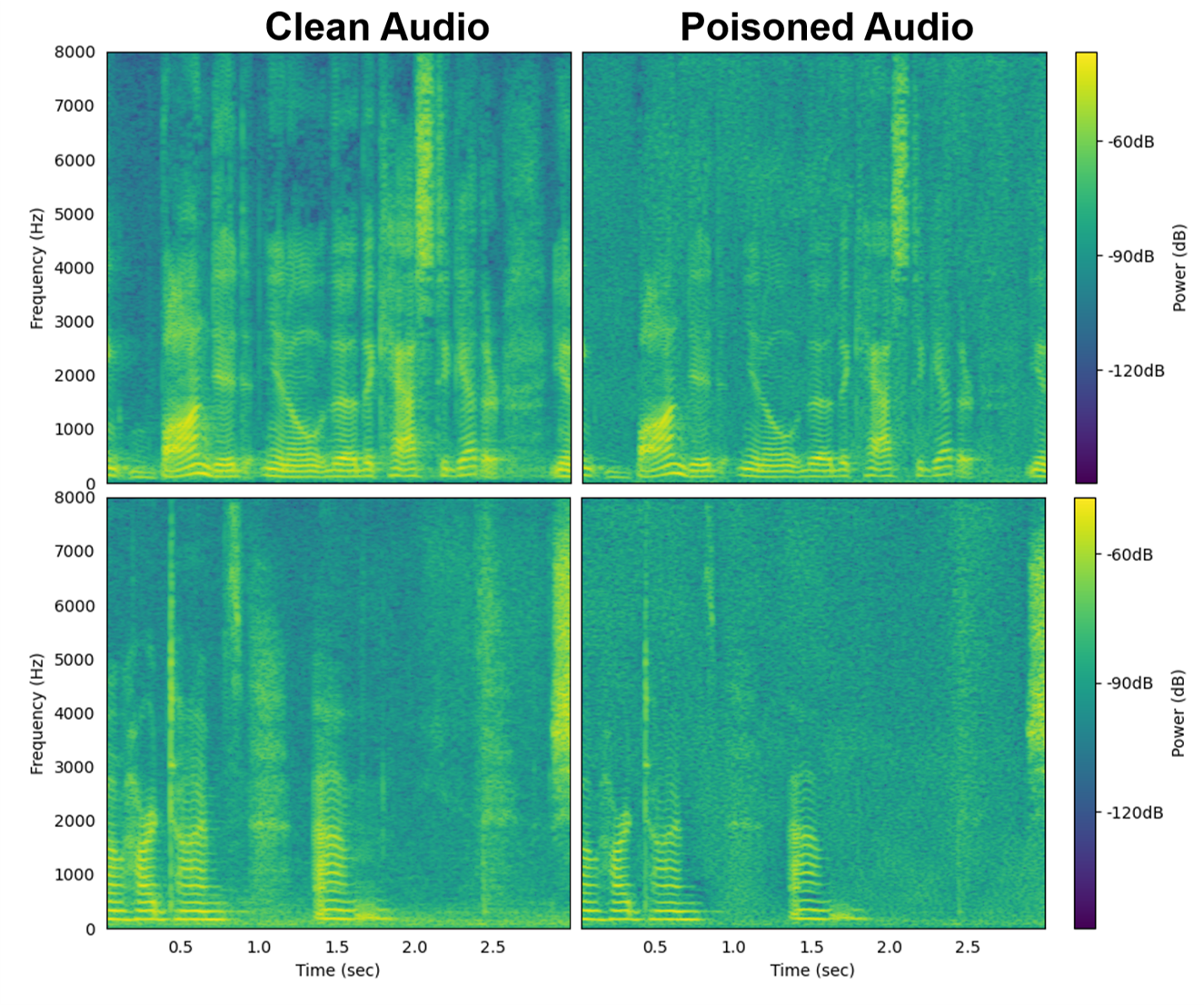}
	\vspace{-1em}
	\caption{Spectrograms of the benign and poisoned voice examples generated by our WaveFuzz. }\label{SP2}
	\vspace{-1em}
\end{figure} 
 \begin{figure}[t]
	\centering
	\includegraphics[width=0.97\linewidth]{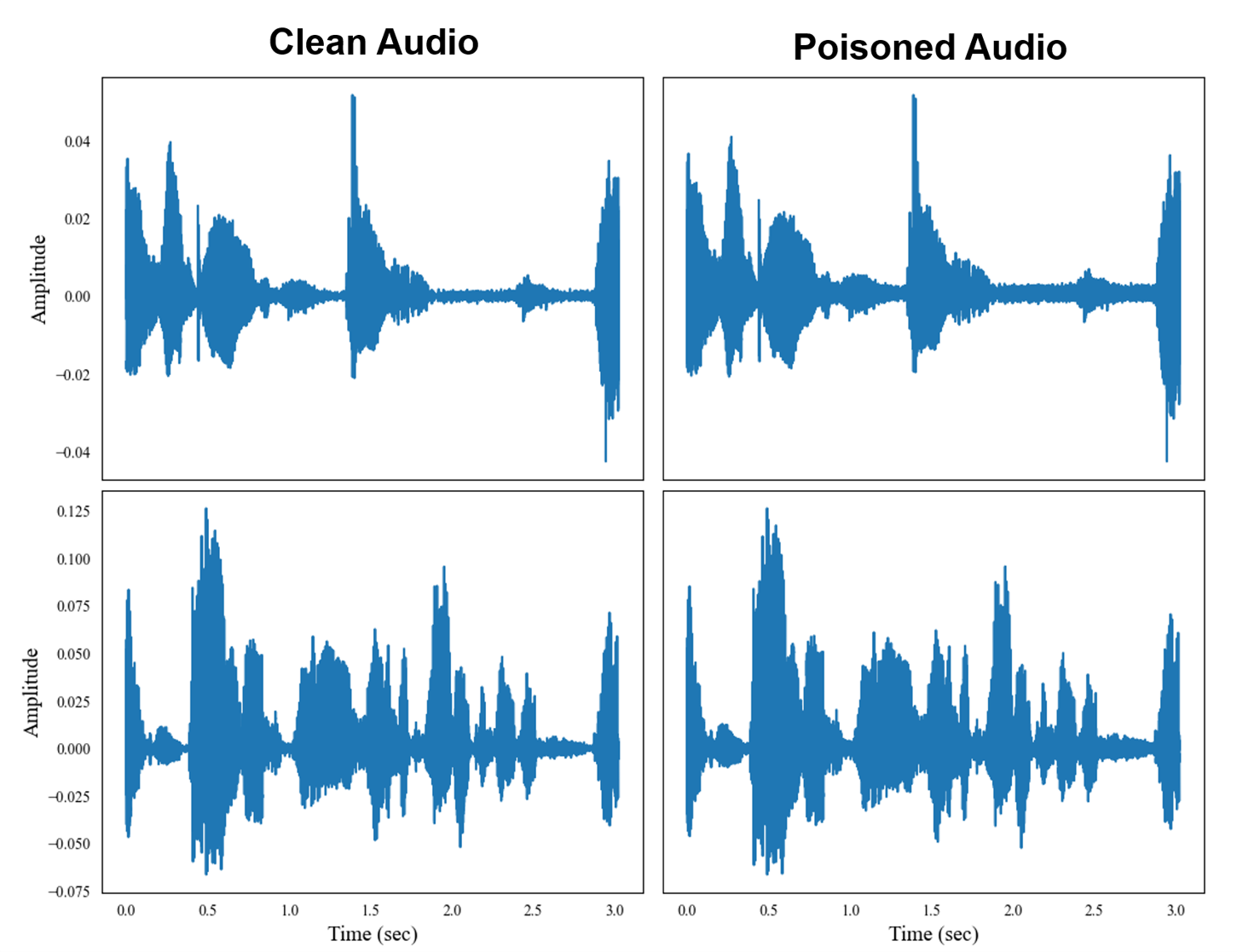}
	\vspace{-1em}
	\caption{Waves of the benign and poisoned voice examples generated by our WaveFuzz. }\label{SP1}
\end{figure}

\begin{table*}[t]
\centering

\resizebox{0.96\linewidth}{!}{
\setlength{\tabcolsep}{1.30mm}{
\begin{tabular}{clcccclcccc} 
\toprule
\multirow{2}{*}{TASK} & \multirow{2}{*}{} & \multicolumn{4}{c}{Poisoned audio}         & \multicolumn{1}{c}{} & \multicolumn{4}{c}{benign audio}          \\ 
\cmidrule{3-6}\cmidrule{8-11}
                      &                   & Normal(\%) & Noise A(\%) & Noise B(\%) & Lable Acc.(\%) &                      & Normal(\%) & Noise A(\%) & Noise B(\%) & Label Acc.(\%)  \\ 
\midrule
SCR                   &                    & 59.50 & 33.00 & 7.50  & 90.50 &  & 94.50 & 5.50  & 0.00 & 96.00 \\
\midrule
SR                    &                    & 69.00 & 27.00 & 4.00  & 88.50 &  & 90.75 & 8.00  & 1.25 & 93.50          \\
\bottomrule
\end{tabular}}}
\caption{Results of the human perception evaluation on poisoned audio generated by WaveFuzz and the corresponding benign audio.}\label{TA:HU}
\end{table*}

\subsection{Attack Performance}
To fully evaluate the performance of our WavFuzz, we target two classical  intelligence audio systems (\ie, SR and SCR). Here, the hackers use the poisoned voice data to fine-tune their pre-trained models or train models from scratch. The attack results are reported in Table~\ref{TA:PE1} and Table~\ref{TA:PE2}. We make the following key observations.

\textbf{Performance Degradation.} 
Based on the results, we find that all the models experienced some degradation in accuracy, implying that the poisoned voice data reduces the expected performance of the malicious models. The performance of the models is severely compromised, especially when the hackers only use the poisoned voice data to fine-tune their models. For example, we only use the 5\% poison data points to fine-tune the models, and the precisions of the malicious models are decreased by  24.55\%, 10.64\%, and 9.75\%, respectively. Moreover, for training the models from scratch, our WaveFuzz is still effective to hinder the models from learning useful information from the protected voice data. Interestingly, the average model accuracy drops by 3.87\% even when only 0.5\% of the samples were poisoned. We also notice the abnormal results from Resnet-18, where only an DAcc. of 2.48 on an PR of 10\%. The effect we attribute to the poor ability of the model where the model accuracy is only 79.07\%.   


\begin{figure}[t]
	\centering
	\includegraphics[width=1\linewidth]{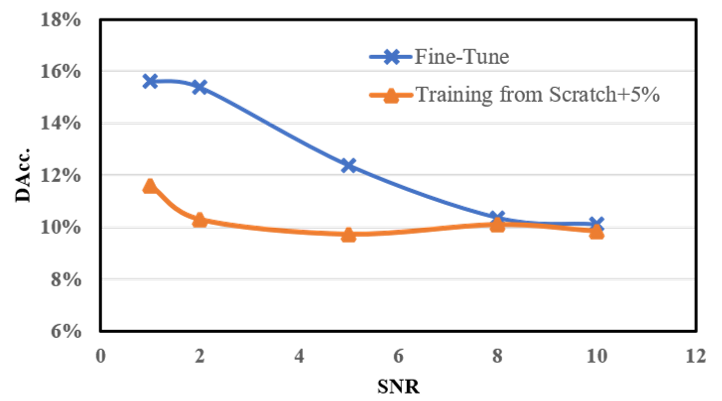}
	\vspace{-1.8em}
	\caption{The DAcc. under different SNR on CNN, where we manual add the DAcc. from training from scratch by 5\% to draw.}\label{fig:snr}
	\vspace{-1em}
\end{figure}



\textbf{DAcc. vs PR.} 
To validate the effectiveness of our WaveFuzz, we conduct the poisoning attack under different PRs (\eg, 10\%, 5\%, 1\%, and 0.5\%). A high PR value means that more protected voice samples are collected and used them to fine-tune or train the models.  Based on these experimental results, we find a consistent conclusion that the more poisoned samples used for training, the more severely the performance of the model suffers.

\textbf{Transferability.} 
The malicious models we use adopt two feature extraction algorithms (\ie, MFCC and FFT). Our WaveFuzz still mitigates privacy leakage since all the model performances are damaged. There is no doubt that the performance of WaveFuzz will be decreased when it encounters some feature extraction algorithms different from MFCC. 
To further improve the generalization ability of WaveFuzz, we plan to integrate various feature extraction algorithms to disturb the benign audio example as future work.

\textbf{Visualization.} 
Fig.~\ref{SP2} shows the spectrograms of the benign voice and poisoned voice examples. First, the two spectrograms are different, proving that the poisoned voice features are confused. Second, though there are apparent differences between the two spectrograms, we cannot visually judge which is wrong since the spectrogram is not intuitionistic.
Fig.~\ref{SP1} shows the waveforms of the benign voice and poisoned voice examples. The two waveforms are close and indistinguishable, indicating that the poisoned voice data can successfully transfer the information to others, and they do not feel abnormal. 



\begin{table}[t]
\centering
\resizebox{0.95\linewidth}{!}{
\setlength{\tabcolsep}{3.30mm}{

\begin{tabular}{ccccc} 
\toprule
\multirow{2}{*}{\begin{tabular}[c]{@{}c@{}}\\PR\end{tabular}} & \multicolumn{2}{c}{WaveFuzz} & \multicolumn{2}{c}{Random noise}  \\ 
\cmidrule(r){2-3}\cmidrule(r){4-5}
                                                              & DAcc.(\%) & SNR              & DAcc.(\%) & SNR                   \\ 
\midrule
\multicolumn{5}{c}{Hacker's Goal: Fine-tuning CNN}                                                                                      \\ 
\midrule
10\%                                                          & 14.37     & 7.89             & 3.62      & 7.91                  \\
5\%                                                           & 9.75      & 7.90             & 2.75      & 7.68                  \\
1\%                                                           & 7.37      & 7.73             & 2.12      & 7.64                  \\
0.50\%                                                        & 5.75      & 7.69             & 1.87      & 7.73                  \\
\midrule
\multicolumn{5}{c}{Hacker's Goal: Training CNN from Scratch}                                                                                      \\ 
\midrule
10\%                                                          & 6.62     & 6.89             & 2.26      & 7.02                  \\
5\%                                                           & 5.45      & 7.45             & 0.62      & 7.24                  \\
1\%                                                           & 5.35      & 7.26             & 0.12      & 7.51                  \\
0.50\%                                                        & 5.23      & 7.76             & 0.77      & 7.69                  \\
\midrule

\multicolumn{5}{c}{Hacker's Goal: Training VggVox from Scratch} \\
\midrule
10\%                                                          & 7.28     & 10.14             & 1.23      &10                  \\
5\%                                                           & 5.23      & 10.28             & -2.77      & 10                  \\
1\%                                                           & 4.00      & 10.23            & 0.01      & 10                  \\
0.50\%                                                        & 3.39      & 10.05             & 0.31      & 10                  \\

\bottomrule
\end{tabular}}}
\caption{The comparison of accuracy decline caused by our WaveFuzz and random noise. We fine-tune CNN, and train CNN and VggVox from scratch.}\label{TA:RE}
\vspace{-1em}
\end{table}

\subsection{Human Study}
We conduct a human study to evaluate the imperceptibility of the corrupted samples. In this experiment, the volunteers are asked to do two studies, where we survey 20 volunteers aged from 18 to 26, including nine males and eleven females. In the first study, given an voice example, the volunteers need to label the example, where they give the content of example for SCR task or judge who is the speaker from three candidates for SR task. We use Label Acc. to represent the rate of the poisoned samples understood by the users. If the label given by the user is the same as the original label, we believe that the current poisoned example can be successfully understood by the user.  
In the second study, the volunteers listen to voice examples and give their views, \ie, ``normal'', ``noise but acceptable'' (NoiseA), and ``noise and unacceptable'' (NoiseB). Specifically, we pick ten pairs of clean and poisoned voice examples for each audio task, where we use the clean voice examples as the reference. Table~\ref{TA:HU} presents the results of the experiments on human study, showing that our poisoned voice can successfully transfer the semantic meaning to humans.

\subsection{Compare with Baseline and Trade-off}
Here, we discuss that our WaveFuzz can damage the model performance, not caused by adding some noises.   
We report the results only adding random noise (gaussian noise) as the reference. 
Table~\ref{TA:RE} shows the DAcc. caused by WaveFuzz and random noise under the same level of PR and SNR . The DAcc. shows that our WaveFuzz will cause more decline in the model performance, where the average values of the differences are 6.49\%, 4.77\%, and 5.28\% for fine-tuning CNN, and training CNN and VggVox from scratch respectively. Besides, the difference between DAcc.s from WaveFuzz and random noise increases with the poisoning rate, \eg, 3.88\% (PR=0.5\%) to 10.75\% (PR=10\%) for fine-tuning,  3.77\% (PR=0.5\%) to 5.21\% (PR=10\%) for training from scratch.

Besides, we also conduct an experiment to analyze the trade-off between the DAcc. and the SNR (see Fig.~\ref{fig:snr}), where the targeted model is the CNN. It is clear that the DAcc. decreases as the SNR increases, whereas a low SNR indicates more noise added in the clean example. The high SNR or slight noise limits us from fully interfering with the sample, so the feature distance between the original example and the poisoned example is not huge enough to confuse the malicious model. Hence, there is a trade-off between the attack performance and stealthiness.


\subsection{WaveFuzz against Pre-Trained Models}
We evaluate whether our poisoned voice data can confuse the pre-trained automatic recognition systems. ECAPA-TDNN~\cite{DesplanquesTD20}, an SR model, achieves the state-of-the-art performance. DeepSpeech~\cite{hannun2014deep} is a popular automatic speech recognition (ASR) model, and Alibaba\footnote{https://ai.aliyun.com/nls/asr} is an ASR online service.
Malicious entities may extract private information from the voice to analyze users' behaviors, which will lead to severe problems of privacy leakage. We generate the poisoned voice by our WaveFuzz and use it to ``attack'' the pre-trained models. If the models returns the output that is different from the label of the corresponding benign voice, we think the malicious model is hard to extract the information contained in the voice. 
The results are shown in Table~\ref{at}, where we calculate accuracies on the benign and poisoned voice data. Particularly, we use word error rate (WER) for ASR task.  
Based on these results, we observe that the protected voice could not be correctly recognized by these recognition systems, meaning that the poisoned voice data is hard to by analyzed by the pre-trained models. 

\begin{table}[t]
\centering
\resizebox{0.95\linewidth}{!}{
\setlength{\tabcolsep}{1.30mm}{
\begin{tabular}{cccccc} 
\toprule
Task & Model      & \begin{tabular}[c]{@{}c@{}}Normal \\Acc.(\%)\end{tabular} & \begin{tabular}[c]{@{}c@{}}Poisoned \\Acc.(\%)\end{tabular} & SNR & $\epsilon$  \\ 
\midrule
\multirow{2}{*}{SR}                                     & VggVox     & 82.29                                                      & 66.15                                                      &10.27     &   1000         \\
                                                        & ECAPA-TDNN &  96.62                                                         &  54.75                                                         &  4.03   & 1700            \\ 
\midrule
\multirow{2}{*}{SCR}                                    & Attention  & 93.28                                                      & 60.12                                                      &  3.88   &  1400           \\
                                                        & CNN       & 91.28                                                          &  58.24                                                         &  7.89    & 1600            \\ 
\midrule
\midrule
Task & Model      & \begin{tabular}[c]{@{}c@{}}Normal \\WER(\%)\end{tabular} & \begin{tabular}[c]{@{}c@{}}Poisoned \\WER(\%)\end{tabular} & SNR & $\epsilon$  \\ 
\midrule

\multirow{2}{*}{ASR}                                    & DeepSpeech      &   23.24                                                        &121.12
                                                           &  3.66   &    3500         \\
                                                        & Alibaba & 12.2                                                        &72.21                                                           & 3.54    & 3500            \\
\bottomrule
\end{tabular}}}
\caption{Classification accuracies of pre-trained models on the clean and poisoned voice data, respectively.}\label{at}
\vspace{-1em}
\end{table}

\section{Discussion}
Previously, we evaluated our WaveFuzz on SR models with the input speech containing short sentences ($\sim$5 words) and SCR models for single-word command.
In this section, we discuss the possibility of preventing more sophisticated automatic speech recognition (ASR), \eg, Kaldi~\footnote{http://www.kaldi-asr.org/}, from stealing useful information from long sentences. Here, we use the Kaldi, a classic ASR model, to discuss the possibility. When PR=0.5\%, the word error rate (WER) is decreased by  0.29\%, where the model generalization ability is improved. WER increases as PR further increases. For example,  the word error rate (WER) increases by 0.42\% when PR=10\%. The results show that our WaveFuzz is not effective enough against Kaldi. The reason is that the prediction model of Kaldi is robust to poisoned features and complicated. 
We plan to design more effective methods to avoid various information in voice data extracted by various audio models as future work. 
\section{Conclusion}
In this paper, we proposed a clean label poisoning attack, WaveFuzz, which can mitigate the risk of users' voice data being misused  and maintain its messaging functionality. Specifically, we generate the protected voice data by maximizing the feature distance between the original voice data and poisoned voice data. We test two classic intelligence audio systems. The models can not steal information from the users' voice data and suffer from the punishment (\ie, performance decline). Moreover, protected voice data can avoid being analyzed by pre-trained recognition systems, fooling these malicious systems. We make a human study to show that poisoned voice data can be understood by other users, indicating that its messaging function is normal. Consequently, WaveFuzz is a simple and effective baseline for mitigating privacy leakage of the user's voice. We think our work sheds new light on ameliorating privacy issues. 


\clearpage

\bibliographystyle{named}
\bibliography{icme2022template}

\begin{thebibliography}{}

\bibitem[\protect\citeauthoryear{Aghakhani \bgroup \em et al.\egroup
  }{2020}]{aghakhani2020venomave}
Hojjat Aghakhani, Lea Sch{\"o}nherr, Thorsten Eisenhofer, Dorothea Kolossa,
  Thorsten Holz, Christopher Kruegel, and Giovanni Vigna.
\newblock Venomave: Targeted poisoning against speech recognition.
\newblock {\em arXiv preprint arXiv:2010.10682}, 2020.

\bibitem[\protect\citeauthoryear{Aghakhani \bgroup \em et al.\egroup
  }{2021}]{aghakhani2021bullseye}
Hojjat Aghakhani, Dongyu Meng, Yu-Xiang Wang, Christopher Kruegel, and Giovanni
  Vigna.
\newblock Bullseye polytope: A scalable clean-label poisoning attack with
  improved transferability.
\newblock In {\em Proc. of {IEE} EuroS\&P}, pages 159--178, 2021.

\bibitem[\protect\citeauthoryear{Arik \bgroup \em et al.\egroup
  }{2017}]{ArikKCHGFPC17}
Sercan~{\"{O}}mer Arik, Markus Kliegl, Rewon Child, Joel Hestness, Andrew
  Gibiansky, Christopher Fougner, Ryan Prenger, and Adam Coates.
\newblock Convolutional recurrent neural networks for small-footprint keyword
  spotting.
\newblock In {\em Proc. of Interspeech}, pages 1606--1610, 2017.

\bibitem[\protect\citeauthoryear{Chung \bgroup \em et al.\egroup
  }{2018}]{abs-1806-05622}
Joon~Son Chung, Arsha Nagrani, and Andrew Zisserman.
\newblock Voxceleb2: Deep speaker recognition.
\newblock {\em CoRR}, 2018.

\bibitem[\protect\citeauthoryear{de Andrade \bgroup \em et al.\egroup
  }{2018}]{deandrade2018neural}
Douglas~Coimbra de~Andrade, Sabato Leo, Martin Loesener Da~Silva Viana, and
  Christoph Bernkopf.
\newblock A neural attention model for speech command recognition, 2018.

\bibitem[\protect\citeauthoryear{Desplanques \bgroup \em et al.\egroup
  }{2020}]{DesplanquesTD20}
Brecht Desplanques, Jenthe Thienpondt, and Kris Demuynck.
\newblock {ECAPA-TDNN:} emphasized channel attention, propagation and
  aggregation in {TDNN} based speaker verification.
\newblock In {\em Proc. of Interspeech}, pages 3830--3834, 2020.

\bibitem[\protect\citeauthoryear{Devaux \bgroup \em et al.\egroup
  }{2009}]{DBLP:conf/mva/DevauxPPC09}
Alexandre Devaux, Nicolas Paparoditis, Fr{\'{e}}d{\'{e}}ric Precioso, and
  Bertrand Cannelle.
\newblock Face blurring for privacy in street-level geoviewers combining face,
  body and skin detectors.
\newblock In {\em Proc. of {MVA}}, pages 86--89, 2009.

\bibitem[\protect\citeauthoryear{Geiping \bgroup \em et al.\egroup
  }{2020}]{GeipingFHCT0G21}
Jonas Geiping, Liam~H. Fowl, W.~Ronny Huang, Wojciech Czaja, Gavin Taylor,
  Michael Moeller, and Tom Goldstein.
\newblock Witches' brew: Industrial scale data poisoning via gradient matching.
\newblock In {\em Proc. of {ICLR}}, 2020.

\bibitem[\protect\citeauthoryear{Gu \bgroup \em et al.\egroup
  }{2017}]{gu2017badnets}
Tianyu Gu, Brendan Dolan-Gavitt, and Siddharth Garg.
\newblock Badnets: Identifying vulnerabilities in the machine learning model
  supply chain.
\newblock {\em arXiv preprint arXiv:1708.06733}, 2017.

\bibitem[\protect\citeauthoryear{Hannun \bgroup \em et al.\egroup
  }{2014}]{hannun2014deep}
Awni Hannun, Carl Case, Jared Casper, Bryan Catanzaro, Greg Diamos, Erich
  Elsen, Ryan Prenger, Sanjeev Satheesh, Shubho Sengupta, Adam Coates, et~al.
\newblock Deep speech: Scaling up end-to-end speech recognition.
\newblock {\em arXiv preprint arXiv:1412.5567}, 2014.

\bibitem[\protect\citeauthoryear{Hermansky}{1990}]{hermansky1990perceptual}
Hynek Hermansky.
\newblock Perceptual linear predictive (plp) analysis of speech.
\newblock {\em the Journal of the Acoustical Society of America}, pages
  1738--1752, 1990.

\bibitem[\protect\citeauthoryear{Huang \bgroup \em et al.\egroup
  }{2020}]{huang2020metapoison}
W~Ronny Huang, Jonas Geiping, Liam Fowl, Gavin Taylor, and Tom Goldstein.
\newblock Metapoison: Practical general-purpose clean-label data poisoning.
\newblock {\em arXiv preprint arXiv:2004.00225}, 2020.

\bibitem[\protect\citeauthoryear{Itakura}{1975}]{itakura1975line}
Fumitada Itakura.
\newblock Line spectrum representation of linear predictor coefficients of
  speech signals.
\newblock {\em The Journal of the Acoustical Society of America}, pages
  S35--S35, 1975.

\bibitem[\protect\citeauthoryear{Jagielski \bgroup \em et al.\egroup
  }{2018}]{jagielski2018manipulating}
Matthew Jagielski, Alina Oprea, Battista Biggio, Chang Liu, Cristina
  Nita-Rotaru, and Bo~Li.
\newblock Manipulating machine learning: Poisoning attacks and countermeasures
  for regression learning.
\newblock In {\em Proc. of {IEEE} {S\&P}}, pages 19--35, 2018.

\bibitem[\protect\citeauthoryear{Kingma and Ba}{2014}]{kingma2014adam}
Diederik~P Kingma and Jimmy Ba.
\newblock Adam: A method for stochastic optimization.
\newblock {\em arXiv preprint arXiv:1412.6980}, 2014.

\bibitem[\protect\citeauthoryear{Liu \bgroup \em et al.\egroup
  }{2017}]{DBLP:journals/scn/LiuZY17}
Yujia Liu, Weiming Zhang, and Nenghai Yu.
\newblock Protecting privacy in shared photos via adversarial examples based
  stealth.
\newblock {\em Secur. Commun. Networks}, 2017:1897438:1--1897438:15, 2017.

\bibitem[\protect\citeauthoryear{Muda \bgroup \em et al.\egroup
  }{2010}]{muda2010voice}
Lindasalwa Muda, Mumtaj Begam, and Irraivan Elamvazuthi.
\newblock Voice recognition algorithms using mel frequency cepstral coefficient
  (mfcc) and dynamic time warping (dtw) techniques.
\newblock {\em arXiv preprint arXiv:1003.4083}, 2010.

\bibitem[\protect\citeauthoryear{Mu{\~n}oz-Gonz{\'a}lez \bgroup \em et
  al.\egroup }{2017}]{munoz2017towards}
Luis Mu{\~n}oz-Gonz{\'a}lez, Battista Biggio, Ambra Demontis, Andrea Paudice,
  Vasin Wongrassamee, Emil~C Lupu, and Fabio Roli.
\newblock Towards poisoning of deep learning algorithms with back-gradient
  optimization.
\newblock In {\em Proc. of AISec}, pages 27--38, 2017.

\bibitem[\protect\citeauthoryear{Nelson \bgroup \em et al.\egroup
  }{2008}]{nelson2008exploiting}
Blaine Nelson, Marco Barreno, Fuching~Jack Chi, Anthony~D Joseph, Benjamin~IP
  Rubinstein, Udam Saini, Charles Sutton, J~Doug Tygar, and Kai Xia.
\newblock Exploiting machine learning to subvert your spam filter.
\newblock {\em LEET}, pages 1--9, 2008.

\bibitem[\protect\citeauthoryear{Oh \bgroup \em et al.\egroup
  }{2017}]{DBLP:conf/iccv/OhFS17}
Seong~Joon Oh, Mario Fritz, and Bernt Schiele.
\newblock Adversarial image perturbation for privacy protection {A} game theory
  perspective.
\newblock In {\em Proc. of {IEEE} {ICCV}}, pages 1491--1500, 2017.

\bibitem[\protect\citeauthoryear{Szegedy \bgroup \em et al.\egroup
  }{2014}]{DBLP:journals/corr/SzegedyZSBEGF13}
Christian Szegedy, Wojciech Zaremba, Ilya Sutskever, Joan Bruna, Dumitru Erhan,
  Ian~J. Goodfellow, and Rob Fergus.
\newblock Intriguing properties of neural networks.
\newblock In {\em Proc. of {ICLR}}, 2014.

\bibitem[\protect\citeauthoryear{Wilber \bgroup \em et al.\egroup
  }{2016}]{DBLP:conf/wacv/WilberSB16}
Michael~J. Wilber, Vitaly Shmatikov, and Serge~J. Belongie.
\newblock Can we still avoid automatic face detection?
\newblock In {\em Proc. of {IEEE} {WACV}}, pages 1--9, 2016.

\bibitem[\protect\citeauthoryear{Xiao \bgroup \em et al.\egroup
  }{2012}]{xiao2012adversarial}
Han Xiao, Huang Xiao, and Claudia Eckert.
\newblock Adversarial label flips attack on support vector machines.
\newblock In {\em Proc. of {ECAI}}, pages 870--875. 2012.

\bibitem[\protect\citeauthoryear{Xiao \bgroup \em et al.\egroup
  }{2015}]{xiao2015feature}
Huang Xiao, Battista Biggio, Gavin Brown, Giorgio Fumera, Claudia Eckert, and
  Fabio Roli.
\newblock Is feature selection secure against training data poisoning?
\newblock In {\em Proc. of {ICML}}, pages 1689--1698, 2015.

\bibitem[\protect\citeauthoryear{Zhu \bgroup \em et al.\egroup
  }{2019}]{zhu2019transferable}
Chen Zhu, W~Ronny Huang, Hengduo Li, Gavin Taylor, Christoph Studer, and Tom
  Goldstein.
\newblock Transferable clean-label poisoning attacks on deep neural nets.
\newblock In {\em International Conference on Machine Learning}, pages
  7614--7623, 2019.

\end{thebibliography}

\end{document}